# Application of a Relativistic Coupled-Cluster Theory to the Effective Electric Field in YbF

M. Abe*, G. Gopakumar, and M. Hada
*Tokyo Metropolitan University, 1-1, Minami-Osawa, Hachioji-city, Tokyo 192-0397, Japan*
*JST, CREST, 4-1-8 Honcho, Kawaguchi, Saitama 332-0012, Japan*

B. P. Das
*Indian Institute of Astrophysics, Bangalore 560 034, India*

H. Tatewaki
*Graduate School of Natural Sciences, Nagoya City University, Nagoya, Aichi 467-8501, Japan*

D. Mukherjee
*Raman Center of Atomic, Molecular and Optical Sciences, IACS, Kolkata 700 032, India*

An accurate determination of the effective electric field ($E_{\text{eff}}$) in YbF is important, as it can be combined with the results of future experiments to give an improved new limit for the electric dipole moment of the electron. We report a relativistic coupled-cluster calculation of this quantity in which all the core electrons were excited. It surpasses the approximations made in the previous reported calculations. We obtain a value of 23.1 GV/cm for $E_{\text{eff}}$ in YbF with an estimated error of less than 10%. The crucial roles of the basis sets and the core excitations in our work are discussed.

The electric dipole moment (EDM) of a non-degenerate system arises from violations of both the parity (P) and the time-reversal (T) symmetries [1]. T violation implies charge parity (CP) violation via CPT theorem [2]. In general, CP violation is a necessary condition for the existence of the EDMs of physical systems, and, in particular, atoms and molecules.

Paramagnetic atoms and molecules are sensitive to the EDM of the electron (eEDM) [3], which is an important probe of the physics beyond the standard model [4]. The eEDM arising from CP violation could also be related to the matter–antimatter asymmetry in the universe [5]. A number of studies using atoms have been performed during the past few decades to extract an upper limit for the eEDM [6]. In general, for heavy polar molecules, the effective electric field experienced by an electron ($E_{\text{eff}}$) obtained from relativistic molecular calculations can be several orders of magnitude larger than that in atoms [7]. Therefore, the experimental observable (i.e., the shift in energy because of the interaction of the electric field with the eEDM) is also several orders of magnitude larger. Owing to the high sensitivity of the eEDM in molecules, there has been a considerable increase in interest in this field during the past decade [8]. Important experimental results have been reported on YbF [9] and ThO [10]. The upper limits of the eEDM are $10.5 \times 10^{-28}$ e cm in YbF and $8.7 \times 10^{-29}$ e cm in ThO. Improved results for both these molecules are expected in the next few years [10, 11]. The abovementioned limit for the eEDM from YbF was estimated using the available calculated values of $E_{\text{eff}}$. These calculations were based on semi-empirical [12], quasi-relativistic [13, 14], four-component Dirac–Fock [15], four-component many-body perturbation theory (MBPT) [16], and four-component relativistic configuration interaction (CI) methods [17].

The aim of the present work is to calculate $E_{\text{eff}}$ in YbF using a rigorous relativistic many-body method, which is more accurate than the methods used in the previous calculations. The method we have chosen is the four-component relativistic coupled-cluster (RCC) method, which is arguably the current gold standard for calculating the electronic structure of heavy atoms and diatomic molecules [18].

The electron EDM interaction Hamiltonian in a molecule can be written as [19]

$$\hat{H}_{eEDM} = -d_e \sum_{i}^{N_e} \beta \boldsymbol{\sigma}_i \cdot \mathbf{E}_{\text{int}}. \quad (1)$$

Here, $d_e$ is the eEDM of an electron, $\beta$ is one of the Dirac matrices, and $\boldsymbol{\sigma}$ are the Pauli spin matrices. $i$ is the index of summation labelling for electrons and $N_e$ is the total number of electrons. $\mathbf{E}_{\text{int}}$ is the electric field acting on an electron in a molecule. The quantity that is of experimental interest in the search for the eEDM is an energy shift ($\Delta E$) of a particular state owing to the interaction Hamiltonian given in Eq. (1). This can be expressed as

$$\Delta E = \langle \Psi | \hat{H}_{eEDM} | \Psi \rangle = -d_e \sum_i^{N_e} \langle \Psi | \beta \sigma_i \cdot \mathbf{E}_{int} | \Psi \rangle \equiv -d_e E_{eff} \quad (2)$$

Here $|\Psi\rangle$ represents the wave function of a molecular state built from single particle four-component Dirac spinors. From Eq. (2), it follows that

$$E_{eff} = \sum_i^{N_e} \langle \Psi | \beta \sigma_i \cdot \mathbf{E}_{int} | \Psi \rangle = \sum_i^{N_e} \langle \Psi | \beta \sigma_i \cdot (\mathbf{E}_A + \mathbf{E}_B + \mathbf{E}_e) | \Psi \rangle$$
$$. \quad (3)$$

The internal electric field in Eq. (3) consists of contributions from the two nuclei ($\mathbf{E}_A$ and $\mathbf{E}_B$) and the electrons ($\mathbf{E}_e$). It is clear that electronic structure calculations are necessary to obtain $E_{eff}$. However, the evaluation of the electronic term is complicated. Using the relationship [20]

$$-d_e \sum_{i=1}^{N_{elec}} \beta \sigma \cdot \mathbf{E}_{int} = \left[ -\frac{d_e}{e} \beta \sigma \cdot \nabla, \hat{H}_0 \right] + 2ic \frac{d_e}{e} \sum_{i=1}^{N_{elec}} \beta \gamma_5 p^2, \quad (4)$$

Eq.(3) can be rewritten as

$$E_{eff} = 2i \frac{c}{e} \sum_i^{N_e} \langle \Psi | \sum_i^{N_{elec}} \beta \gamma_5 p_i^2 | \Psi \rangle, \quad (5)$$

where $\gamma_5$ is the product of the four Dirac matrices. Eq. (5) represents the expectation value of an effective one body operator and it is therefore much simpler to evaluate Eq. (5) than Eq. (3).

To evaluate $E_{eff}$, we use the ground state wave function of the Dirac–Coulomb (DC) Hamiltonian:

$$\hat{H}_{DC} = \sum_i^{N_{elec}} \left[ c\alpha p_i + \beta c^2 - \sum_A^{N_{nuc}} \frac{Z_A}{|\mathbf{r}_i - \mathbf{R}_A|} \right] + \sum_{i<j}^{N_{elec}} \frac{1}{|\mathbf{r}_i - \mathbf{r}_j|}, \quad (6)$$

Here, $c$ is the speed of light, $p$ is the momentum operator, $\alpha$ collectively represents the Dirac matrices, $\mathbf{r}$ and $\mathbf{R}$ refer to the position vectors of the electrons and the nuclei, respectively, and $Z$ is the charge of a nucleus. The capital letters in the summation are the nuclear indices while the small letters are the electronic indices.

We used the RCC method to obtain the ground state wave function of YbF. The reference state was taken as a single determinant corresponding to an open shell doublet at the Dirac–Fock level. The Dirac–Fock configuration obtained was almost ionic, such as Yb:(6$s$)[1] and F:(2$p$)[6]. The coupled-cluster wave function is given by

$$|\psi_{CCSD}\rangle = \exp(\hat{T}_1 + \hat{T}_2)|\psi_{DF}\rangle, \quad (7)$$

where $\hat{T}_1$ and $\hat{T}_2$ are single (S) and double (D) excitation operators, respectively. The salient features of our relativistic coupled-cluster singles and doubles (RCCSD) method are: (i) It uses the Dirac–Coulomb approximation. (ii) It treats correlation effects to all orders in the residual Coulomb interaction for one and two hole-particle excitations. (iii) It is size extensive unlike the truncated CI [21].

We modified and combined two of the most widely used relativistic codes, REL4D in UTChem [22] and DIRAC08 [23], for our calculations. We used UTChem for the generation of the Dirac–Fock orbitals and the molecular orbital (MO) integral transformations [24]. We developed a suitable computational algorithm in UTChem to evaluate one-electron integrals of the effective EDM Hamiltonian used in Eq. (5). We used the $C_8$ double group symmetry, which is available in UTChem [25] but not in DIRAC. The adaptation of $C_8$ point group drastically reduces the computational costs for both the MO transformation and the RCCSD calculations. Using the MO integrals, the DIRAC08 code was used to obtain the RCCSD amplitudes. In the present work, the expectation value of a normal ordered operator was calculated by considering only the linear terms in the coupled-cluster wave function as they make the most important contributions. We therefore express the expectation value as

$$\langle \psi_{DF} | (1 + T_1^+ + T_2^+) \hat{O}_N (1 + T_1 + T_2) | \psi_{DF} \rangle_C, \quad (8)$$

where the subscript $C$ in Eq. (8) indicates that we calculate all the fully contracted terms. $E_{eff}$ and the molecular electric dipole moment ($DM$) were calculated using Eq. (8), where $\hat{T}_1$ and $\hat{T}_2$ amplitudes were determined from the CCSD amplitude equations, as mentioned earlier. In the framework of coupled-cluster theory, the normal CC method [26] and its variant the lambda CC method [27] are two of the widely used approaches to calculate expectation values. These methods require substantially more computational time than the CC method used in the present work. In addition, our approach captures the dominant correlation effects.

We used Dyall's four-component valence double zeta (DZ), triple zeta (TZ), and quadruple zeta (QZ) basis sets for ytterbium. [28] For fluorine, we used Watanabe's four-component basis sets. [29] In addition, we used some diffuse and polarization functions from the Sapporo basis sets. [30] All of the basis sets were used in the uncontracted form. The QZ basis is the most accurate among the ones considered and its accuracy has been confirmed by Gomes *et al.* [28] The bond length and harmonic frequency they obtained with the QZ basis were 2.0196 Å and 503.2 cm$^{-1}$, respectively. These results are very close to the experimental values, 2.0161 Å and 506.6674 cm$^{-1}$ [31], and also close to the extrapolated values, 2.0174 Å and 507.6 cm$^{-1}$, obtained from the results of DZ, TZ, and QZ basis sets.

We performed two different relativistic CCSD calculations to investigate the importance of core correlation effects. They are referred to as 49e-CCSD and 79e-CCSD. In the first case, 49 electrons were excited, that is, the 3*s*, 3*p*, 3*d*, 2*s*, 2*p* and 1*s* orbitals of Yb and the 1s orbital of F are frozen. In the second case, all 79 electrons of YbF were excited. We performed only a single geometry calculation with the experimental value of bond length of 2.0161 Å.

Table I summarizes our calculated results. At the Dirac–Fock level, the values of $E_{eff}$ and *DM* for the three basis sets are very close. However, there is a basis set dependence in the 49e-CCSD calculation. For the TZ basis set, we obtained a relatively large value for $T_1$ diagnostic (0.0558), which indicates instability of the single-reference calculations with the TZ basis. Gomes *et al.* also encountered the same problem at the TZ basis and discussed its influence on the spectroscopic constants [28]. In our QZ basis calculation, the $T_1$ diagnostic is lower (0.0397) than that for the TZ basis. Therefore, it is reasonable to assume that our QZ basis calculations provide reliable results for $E_{eff}$ and *DM*. The change in the values of *DM* for our two CCSD/QZ (49e-CCSD and 79e-CCSD) calculations was virtually negligible. However, the corresponding change in the case of $E_{eff}$ was significantly larger, underlining the relative importance of core-correlation for this property. Our best result was 23.1 GV/cm for $E_{eff}$ and 3.60 Debye for *DM* obtained from the fully core-correlated calculations with the QZ basis set. The calculated value of *DM* was within 8% of the measured value of this quantity (3.91 Debye). Both $E_{eff}$ and *DM* depend on the mixing of orbitals of opposite parities. Another important property that bears some resemblance to $E_{eff}$ is the hyperfine coupling constant (HFCC). Both these quantities are sensitive to the behavior of the wave function in the region of nucleus. The parallel component of the HFCC ($A_{//}$) was obtained as 6239 and 7913 MHz at the DHF/QZ and 79e-CCSD/QZ levels, respectively. The corresponding experimental value is 7424 MHz [32]. Thus, our best calculation of $A_{//}$ is within 7% of the experimental value.

$E_{eff}$ in YbF has been previously calculated by different methods. The earliest work by Titov *et al.* in 1996 [13] was based on the restricted-active-space self-consistent-field (RASSCF) method with the generalized relativistic effective core potential (GRECP). They obtained $E_{eff}$ = 18.8 GV/cm. This result was later improved to 24.9 GV/cm with an effective operator for core polarization by Mosyagin *et al.* in 1998 [14]. Another study by Kozlov in 1997 [12] based on a semi-empirical method calculated $E_{eff}$ = 26.1 GV/cm. A comparison with the four-component relativistic methods is necessary to assess the accuracy of the abovementioned approximations.

In the framework of the four-component Dirac relativistic method, Parpia [15] calculated $E_{eff}$ = 19.9 GV/cm from the unpaired spinor, and 24.9 GV/cm from all the occupied spinors at the unrestricted DF level in 1998. In the same year, Quiney *et al.* [16] calculated $E_{eff}$ = 24.8 GV/cm at the restricted DF level with the first-order core polarization included by MBPT. In 2006 and 2009, Nayak *et al.* [17] accounted for electron correlation in $E_{eff}$ using the restricted-active-space configuration interaction (RASCI) method at the four-component level. Their best result was $E_{eff}$ = 24.1 GV/cm from a 31-electron correlated calculation in a space of 76 active orbitals. This active space is not sufficiently large. Our relativistic CCSD calculation has the following advantages over the relativistic CI approach used by Nayak *et al.* [17]. (i) CCSD is size extensive unlike the approximate CI used in the latter work. (ii) The size of our QZ basis is substantially larger than that used in Ref. [17]. (iii) We included all the core-correlation effects, while only 31 electrons were excited in the CI calculation. (iv) The number of spinors in our virtual space (293 orbitals) is also much larger than in Ref .[17] (60 orbitals).

Electron correlation increased the value of $E_{eff}$ by about 20% in our work. This trend is similar to the previous correlation calculations, especially where core-polarization effects were included. In conclusion, our most accurate calculation of the effective field gives $E_{eff}$ = 23.1 GV/cm. This leads to an increase in the value of the upper limit of the electron EDM ($11.8 \times 10^{-28}$ e cm) compared to the earlier result ($10.5 \times 10^{-28}$ e cm) for YbF. Taking into consideration the omitted higher-order correlation effects and the deviations of our *DM* and HFCC results from their measured values, we estimate an error of 5–8% of in our calculation of $E_{eff}$. In principle, the method that we used can be applied to any lowest energy state that can be described by a single reference determinant in a given irreducible representation. In the future, we will apply the method to some of the molecular candidates that have been proposed in the context of the search for the electron EDM.

We have provided pertinent details about the basis sets, treatment of correlation effects, analysis of the dominant contributions from the individual terms in $E_{eff}$, and the definition of the HFC operator used in our work in the supplementary material. We acknowledge valuable discussions with E. Hinds, M. Tarbutt, B. Sauer, A. Vutha, A. Titov, A. N. Petrov, L. Skripnikov, M. K. Nayak, T. Fleig, and B. K. Sahoo. This research was supported by JST, CREST. MA, MH and GG thank MEXT for financial support.

TABLE I. Summary of the calculated results of the present work.

| Method/Basis | $T_1$ Diag. | $E_{eff}$ (GV/cm) | $A_{//}$ (MHz) | $DM$[a] (D) |
|---|---|---|---|---|
| DF/DZ | - | 17.9 | - | 3.21 |
| DF/TZ | - | 18.2 | - | 3.21 |
| DF/QZ | - | 18.2 | 6239 | 3.21 |
| 49e-CCSD/DZ | 0.0432 | 21.4 | - | 3.37 |
| 49e-CCSD/TZ | 0.0588 | 21.1 | - | 3.46 |
| 49e-CCSD/QZ | 0.0397 | 22.7 | - | 3.59 |
| 79e-CCSD/QZ | 0.0311 | 23.1 | 7913 | 3.60 |
| Exp. | - | - | 7424[b] | 3.91[c] |

[a]The direction of the dipole moment is taken as the molecular axis from the fluorine to the ytterbium atom.

[b]Ref. [32]

[c]Ref. [31]